# Coding for Parallel Channels: Gallager Bounds and Applications to Repeat-Accumulate Codes

Igal Sason    Idan Goldenberg
Department of Electrical Engineering
Technion - Israel Institute of Technology
Haifa 32000, Israel
Email: {sason@ee, idang@tx}.technion.ac.il

*Abstract*— **This paper is focused on the performance analysis of binary linear block codes (or ensembles) whose transmission takes place over independent and memoryless parallel channels. New upper bounds on the maximum-likelihood (ML) decoding error probability are derived. The framework of the second version of the Duman and Salehi (DS2) bounds is generalized to the case of parallel channels, along with the derivation of optimized tilting measures. The connection between the generalized DS2 and the 1961 Gallager bounds, known previously for a single channel, is revisited for the case of parallel channels. The new bounds are used to obtain improved inner bounds on the attainable channel regions under ML decoding. These improved bounds are applied to ensembles of turbo-like codes, focusing on repeat-accumulate codes and their recent variations.**

## I. INTRODUCTION

Performance analysis of linear block codes whose transmission takes place over parallel channels is of interest since communication over parallel channels models various practical scenarios. In this case the message is partitioned into several disjoint subsets, and the bits in each subset are transmitted over a different channel. This enables one to model transmission over block-fading channels, rate-compatible puncturing of turbo-like codes, incremental redundancy retransmission schemes, cooperative coding, and multi-carrier signaling.

Tight analytical bounds serve as a potent tool for assessing the performance of modern error-correction schemes, both for the case of finite block length and in the asymptotic case where the block length tends to infinity. For a single communication channel, these bounds can be applied in order to obtain a noise threshold which indicates the minimum channel conditions necessary for reliable communication. When generalizing to the scenario of independent parallel channels, this threshold is transformed into a multi-dimensional barrier in terms of the parallel-channel parameters, dividing the space into attainable and non-attainable channel regions.

The performance of code ensembles operating under this setting was recently addressed by Liu et al. [6]; their analysis adapts the 1961 Gallager-Fano bounding technique [4] for communication over parallel channels. The upper bounds on the ML decoding error probability enable to derive achievable regions which ensure reliable communications under ML decoding where the block length of the codes (or ensembles) tend to infinity; these regions depend on the asymptotic distance spectrum of the ensemble of binary linear block codes under consideration (to be defined in Section II).

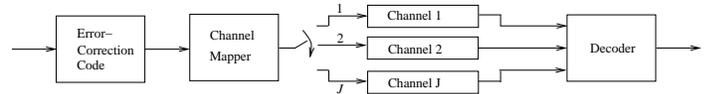

Fig. 1.  System model of parallel channels.

Using a similar approach, we derive a parallel-channel generalization of the DS2 bound introduced in [3], [8]; for the case of parallel channels, we re-examine the well-known relations which exist for the single channel case between this bound and the 1961 Gallager bound (see [2], [8]).

The new bounds are compared with the ones in [6] for parallel Gaussian channels, and show significant improvement over these bounds. They are also exemplified for various ensembles and provide tighter inner bounds on the boundary of the channel regions which are asymptotically attainable under ML decoding. The tightness of these bounds is exemplified for ensembles of accumulate-based codes.

The remainder of the paper is organized as follows. The system model is presented in Section II, as well as preliminary material related to our discussion. In Section III, we derive the improved upper bounds under ML decoding when the transmission takes place over parallel MBIOS channels. Attainable channel regions for ensembles of codes are derived in Section IV. Numerical results are given in Section V. The reader is referred to the full paper version [9] where proofs and further mathematical details are provided.

## II. PRELIMINARIES

### A. System Model

The communication model consists of $J$ statistically independent, memoryless, binary-input and output-symmetric (MBIOS) channels, as shown in Fig. 1. Using a block code $\mathcal{C}$ of size $M = 2^k$, the encoder selects a codeword with equal probability ($\frac{1}{M}$) for transmission. Each codeword consists of $n$ symbols, and the coding rate is defined as $R \triangleq \frac{\log_2 M}{n} = \frac{k}{n}$.

The channel mapper selects for each coded symbol one of $J$ channels through which it is transmitted; the transition probability of the $j$-th channel is $p(y|x;j)$. The received vector is maximum-likelihood (ML) decoded at the receiver where the specific channel mapping is assumed to be known.

Liu et al. [6] introduced the concept of a random channel mapping device which takes a symbol and assigns it to channel $j$ with probability $\alpha_j$, and the assignment is independent of that of other symbols. This approach enables the derivation of an upper bound for the parallel channels which is averaged over all possible channel assignments and calculated in terms of the distance spectrum of the code (or ensemble).

*B. Distance Properties of Ensembles of Turbo-Like Codes*

Bounds on the ML decoding error probability of binary linear block codes or ensembles are often based on their distance properties (see, e.g., [8] and references therein).

Let $[\mathcal{C}(n)]$ be an ensemble of codes of length $n$. We also consider a *sequence of ensembles* $[\mathcal{C}(n_1)], [\mathcal{C}(n_2)], \ldots$ all possessing a common rate $R$. For a given binary linear block code $\mathcal{C}$, let $A_h^{\mathcal{C}}$ (or simply $A_h$) denote the distance spectrum, i.e., the number of codewords of Hamming weight $h$. Referring to ensembles of codes, denote by $A_h^{[\mathcal{C}(n)]}$ the average distance spectrum of the ensemble. We are interested in studying the asymptotic case where $n \to \infty$. To this end, we define the *asymptotic exponent of the distance spectrum* as

$$r^{[\mathcal{C}]}(\delta) \triangleq \lim_{n \to \infty} r^{[\mathcal{C}(n)]}(\delta), \quad r^{[\mathcal{C}(n)]}(\delta) \triangleq \frac{\ln A_h^{[\mathcal{C}(n)]}}{n} \quad (1)$$

where $\delta \triangleq \frac{h}{n}$ is the *normalized distance*.

## III. IMPROVED BOUNDS FOR INDEPENDENT PARALLEL CHANNELS UNDER ML DECODING

*A. The DS2 Bound*

In this section, we present a generalization of the DS2 bound to the case of independent parallel MBIOS channels. For a discussion on the DS2 bound for a single MBIOS channel, the reader is referred to the tutorial paper [8, Chapter 4].

*Theorem 1 [The Generalized DS2 Bound for Parallel Channels]:* Consider the transmission of binary linear block codes (or ensembles) under the scenario described in Section II-A. We have the following results:

- The ML decoding error probability is upper-bounded by

$$P_{\mathrm{e}} \leq \left\{ \sum_{h=0}^{n} A_h \left( \sum_{j=1}^{J} \alpha_j A(\lambda, \rho; j, \psi(\cdot;j)) \right)^h \left( \sum_{j=1}^{J} \alpha_j B(\rho; j, \psi(\cdot;j)) \right)^{n-h} \right\}^{\rho} \quad (2)$$

where $0 \leq \rho \leq 1, \lambda \geq 0$,

$$A(\lambda, \rho; j, \psi(\cdot;j)) \triangleq \sum_y \left( \psi(y;j)^{1-\frac{1}{\rho}} p(y|0;j)^{\frac{1-\lambda\rho}{\rho}} p(y|1;j)^{\lambda} \right)$$

$$B(\rho; j, \psi(\cdot;j)) \triangleq \sum_y \psi(y;j)^{1-\frac{1}{\rho}} p(y|0;j)^{\frac{1}{\rho}}$$

and $\psi(\cdot;j)$, $j = 1, \ldots, J$ are arbitrary probability tilting measures (which are subject to optimization, together with the parameters $\lambda$ and $\rho$). The bound in (2) is referred to as the DS2 bound for parallel channels.

- Consider the partitioning of the code $\mathcal{C}$ to constant Hamming-weight subcodes $\{\mathcal{C}_h\}_{h=1}^n$ where the subcode $\mathcal{C}_h$ consists of all the codewords in $\mathcal{C}$ of Hamming weight $h$ plus the all-zero codeword. Applying the union bound over the subcodes yields

$$P_{\mathrm{e}} \leq \sum_{h=0}^{n} P_{\mathrm{e}|0}(h) \quad (3)$$

where $P_{\mathrm{e}|0}(h)$ is the decoding error probability for the subcode $\mathcal{C}_h$ given that the all-zero codeword is transmitted. Rather than optimizing the tilting measure in the bound given by Eq. (2), we apply the DS2 bound for every subcode and use Eq. (3) to evaluate the overall bound on $P_{\mathrm{e}}$. The key result is that applying the bound in this manner allows us to optimize the $J$ tilting measures for each subcode *separately*. The total number of subcodes does not exceed the block length of the code (or ensemble). Consequently, the use of the union bound does not degrade the related error exponent of the overall bound, but on the other hand, the optimized tilting measures are tailored for each of the constant-Hamming weight subcodes, a process which can only improve the exponential behavior of the resulting bound. The conditional ML decoding error probability of the constant-weight subcode $\mathcal{C}_h$ is therefore upper-bounded by

$$P_{\mathrm{e}|0}(h) \leq (A_h)^{\rho} \left( \sum_{j=1}^{J} \alpha_j A(\lambda, \rho; j, \psi(\cdot;j)) \right)^{h\rho} \left( \sum_{j=1}^{J} \alpha_j B(\rho; j, \psi(\cdot;j)) \right)^{(n-h)\rho} \quad (4)$$

which can be written equivalently in the exponential form

$$P_{\mathrm{e}|0}(h) \leq e^{-n E_\delta^{\mathrm{DS2}}(\lambda, \rho, J, \{\alpha_j\})}$$

where

$$E_\delta^{\mathrm{DS2}}(\lambda, \rho, J, \{\alpha_j\}) \triangleq -\rho r^{\mathcal{C}}(\delta)$$
$$- \rho \delta \ln \left( \sum_{j=1}^{J} \alpha_j A(\lambda, \rho; j, \psi(\cdot;j)) \right)$$
$$- \rho(1-\delta) \ln \left( \sum_{j=1}^{J} \alpha_j B(\rho; j, \psi(\cdot;j)) \right). \quad (5)$$

This exponential form will be useful in Section IV for the discussion on attainable channel regions.

The optimized set of probability tilting measures $\{\psi(\cdot;j)\}_{j=1}^J$ which attains the minimal value of the upper

bound (4) is given (for $j = 1, \ldots, J$) by

$$\psi(y;j) = \beta_j p(y|0;j) \left[1 + k \left(\frac{p(y|1;j)}{p(y|0;j)}\right)^\lambda\right]^\rho. \quad (6)$$

The optimal parameters $k$ and $\beta_j$ $j = 1, \ldots, J$ are related by the following implicit equations

$$k = \frac{\delta}{1-\delta} \frac{\sum_{j=1}^{J} \sum_{y \in \mathcal{Y}} \{\alpha_j C(y;j)\}}{\sum_{j=1}^{J} \sum_{y \in \mathcal{Y}} \left\{\alpha_j C(y;j) \left(\frac{p(y|1;j)}{p(y|0;j)}\right)^\lambda\right\}} \quad (7)$$

where

$$C(y;j) \triangleq \beta_j^{1-\frac{1}{\rho}} p(y|0;j) \left[1 + k \left(\frac{p(y|1;j)}{p(y|0;j)}\right)^\lambda\right]^{\rho-1}$$

$$\beta_j = \left[\sum_{y \in \mathcal{Y}} p(y|0;j) \left(1 + k \left(\frac{p(y|1;j)}{p(y|0;j)}\right)^\lambda\right)^\rho\right]^{-1}. \quad (8)$$

In order to evaluate the bound in (4) for a fixed pair of $\lambda$ and $\rho$, we find the optimized tilting measures in (6) by first assuming an initial vector $(\beta_1, \ldots, \beta_J)$ and then iterating between (7) and (8) until we get a fixed point for these equations. For a fixed $\delta$, we need to optimize numerically the bound in (4) w.r.t. the two parameters $\lambda$ and $\rho$.

### B. The 1961 Gallager Bound

In this section we discuss the 1961 Gallager bound for parallel MBIOS channels.

*Theorem 2 [The 1961 Gallager Bound for Parallel Channels]:* Consider the transmission of binary linear block codes (or ensembles) under the scenario described in Section II-A. We have the following results:

- The 1961 Gallager bound for parallel channels is given by (see [6])

$$P_{\text{e}} \leq 2^{h(\rho)} \left\{ \sum_{h=1}^{n} A_h \left[\sum_{j=1}^{J} \alpha_j Z(r;j)\right]^h \left[\sum_{j=1}^{J} \alpha_j G(r;j)\right]^{n-h} \right\}^\rho \left\{\sum_{j=1}^{J} \alpha_j G(s;j)\right\}^{n(1-\rho)}$$

where $r \leq 0, \quad s \geq 0$,

$$G(r;j) \triangleq \sum_y p(y|0;j)^{1-r} f(y;j)^r$$

$$Z(r;j) \triangleq \sum_y [p(y|0;j)p(y|1;j)]^{\frac{1-r}{2}} f(y;j)^r$$

$$\rho \triangleq \frac{s}{s-r}, \quad 0 \leq \rho \leq 1$$

and $f(\cdot;j)$ is an arbitrary tilting measure, constrained to be non-negative and even (i.e., $f(y;j) = f(-y;j)$).

- By partitioning the code into constant Hamming-weight subcodes and using the union bound as in (3), we obtain optimized tilting measures for the 1961 Gallager bound. The optimal choice for the set of functions $\{f(\cdot;j)\}$ is given by

$$f(y;j) = \left\{ \frac{(1-c)\left(p(y|0;j)^{\frac{1-s(1-\rho^{-1})}{2}} - p(y|1;j)^{\frac{1-s(1-\rho^{-1})}{2}}\right)^2}{p(y|0;j)^{1-s} + p(y|1;j)^{1-s}} \right.$$

$$\left. + \frac{2c\left(p(y|0;j)p(y|1;j)\right)^{\frac{1-s(1-\rho^{-1})}{2}}}{p(y|0;j)^{1-s} + p(y|1;j)^{1-s}} \right\}^{\frac{\rho}{s}}, \quad (\rho, s, c) \in [0,1]^3.$$

where the parameters $\rho, s, c$ are optimized numerically so as to get the tightest bounds within this form.

*Discussion.* In [6], simple choices for these functions are used to evaluate the 1961 Gallager bound for parallel channels, rather than the optimized tilting measures. Consequently, the bounds in [6] are looser.

The connection between the DS2 bound and the 1961 Gallager bound for a single MBIOS channel has been explored in [2], [8]. In this case, it was demonstrated that the DS2 bound is tighter than the Gallager bound. At first, one would expect that a similar result would also hold for the general case where the communication takes place over $J$ independent parallel MBIOS channels. It is shown in [9] that this is not necessarily the case. Consequently, we cannot draw a conclusion establishing the superiority of one of these bounds over the other for the case where $J > 1$. We present numerical results which verify this observation. For technical details, the reader is referred to [9].

## IV. INNER BOUNDS ON ATTAINABLE CHANNEL REGIONS

A $J$-tuple of transition probabilities characterizing a parallel channel is said to be an *attainable channel point* with respect to a code ensemble $\mathcal{C}$ if the average ML decoding error probability vanishes as we let the block length tend to infinity. The *attainable channel region* of an ensemble whose transmission takes place over parallel channels is defined as the closure of the set of attainable channel points. Since the exact decoding error probability under ML decoding is in general unknown, we evaluate inner bounds on the attainable channel regions whose calculation is based on upper bounds on the ML decoding error probability.

Our numerical results referring to inner bounds on attainable channel regions are based on the following theorem.

*Theorem 3 [Inner bounds on the attainable channel regions for parallel channels]:* Consider the transmission of a sequence of binary linear block codes (or ensembles) $\{[\mathcal{C}(n)]\}$ takes place over a set of $J$ parallel MBIOS channels. Let

$$\gamma_j \triangleq \sum_y \sqrt{p(y|0;j)p(y|1;j)}, \quad j \in \{1, \ldots, J\}$$

designate the Bhattachryya constants of the channels. Assume that the following conditions hold:

1)
$$\inf_{\delta_0 < \delta \leq 1} E^{\text{DS2}}(\delta) > 0, \quad \forall \delta_0 \in (0,1)$$

where, for $0 < \delta \leq 1$, $E^{\text{DS2}}(\delta)$ is calculated from (5) by maximizing w.r.t. $\lambda$, $\rho$ ($\lambda \geq 0$ and $0 \leq \rho \leq 1$) and the probability tilting measures $\{\psi(\cdot;j)\}_{j=1}^{J}$.

2) The inequality
$$\limsup_{\delta \to 0} \frac{r^{[\mathcal{C}]}(\delta)}{\delta} < -\ln\left(\sum_{j=1}^{J} \alpha_j \gamma_j\right)$$
is satisfied.

3) There exists a sequence $\{D_n\}$ of natural numbers tending to infinity with increasing $n$ so that
$$\limsup_{n \to \infty} \sum_{h=1}^{D_n} A_h^{[\mathcal{C}(n)]} = 0.$$

4) The normalized exponent of the distance spectrum $r^{[\mathcal{C}(n)]}$ converges uniformly in $\delta \in [0,1]$ to its asymptotic limit (1).

Then, the $J$-tuple vector of parameters characterizing these channels lies within the attainable channel region under ML decoding.

*Discussion.* We note that conditions 3 and 4 in Theorem 3 are similar to the last two conditions in [5, Theorem 2.3]. Condition 2 above happens to be a natural generalization of the second condition in [5, Theorem 2.3] to a set of parallel channels. The distinction between [5, Theorem 2.3] which relates to typical-pairs decoding over a single channel and the statement in Theorem 3 for ML decoding over a set of independent parallel channels lies mainly in the first condition of both theorems.

A similar result which involves the generalized 1961 Gallager bound for parallel channels is given in a similar fashion by replacing the first condition with an equivalent relation involving the exponent of the 1961 Gallager bound maximized over its parameters, instead of the error exponent of the DS2 bound.

## V. Performance Bounds for Turbo-Like Ensembles over Parallel Channels

In this section, we exemplify the performance bounds derived in this paper for various ensembles of turbo-like codes whose transmission takes place over parallel Gaussian channels. We also compare the bounds to those introduced in [6], showing the superiority of the new bounds with their optimized tilting measures.

### A. Performance Bounds for Uniformly Interleaved Turbo Codes

Fig. 2 compares upper bounds on the bit error probability of the ensemble of uniformly interleaved turbo codes of rate $R = \frac{1}{3}$ bits per channel use. The encoder consists of two convolutional encoders with polynomials $G(D) = \left[1, \frac{1+D^4}{1+D+D^2+D^3+D^4}\right]$ and the interleaver between them is of length 1000. The transmission is assumed to take place over two (independent) parallel binary-input AWGN channels where each bit is equally likely to be assigned to one of these channels ($\alpha_1 = \alpha_2 = \frac{1}{2}$), and the value of the energy per bit to spectral noise density of the first channel is set to $\left(\frac{E_b}{N_0}\right)_1 = 0$ dB. The parallel-channel version of the union bound shown in the figure (see [9, Appendix C]) is obtained by averaging the union bound over all channel assignments, a similar process by which the improved bounds are obtained. The LMSF bound (yielding the tightest modified Shulman-Feder [10] bound) is a combination of the union bound with the Shulman-Feder bound (see [6], [9] for details on the derivation of this bound), and was presented in [6] as a special case of the 1961 Gallager bound. Clearly, the DS2 and the 1961 Gallager bounds with their *optimized tilting measures* show a remarkable improvement in their tightness over the union and LMSF bounds in [6]; for a bit error probability of $10^{-4}$, the improved bounds exhibit a gain of 0.8 dB over the LMSF bound. The DS2 bound gains very little over the 1961 Gallager bound (about 0.05 dB) at a bit error probability of $10^{-3}$. In spite of the remarkable advantage of the improved bounds over the union and LMSF bounds, computer simulations under (the sub-optimal) iterative Log-MAP decoding with 10 iterations show a gap of about 0.4 dB in favor of the performance under iterative decoding. This indicates that there is still room for further improvement in the tightness of the analytical bounds under ML decoding.

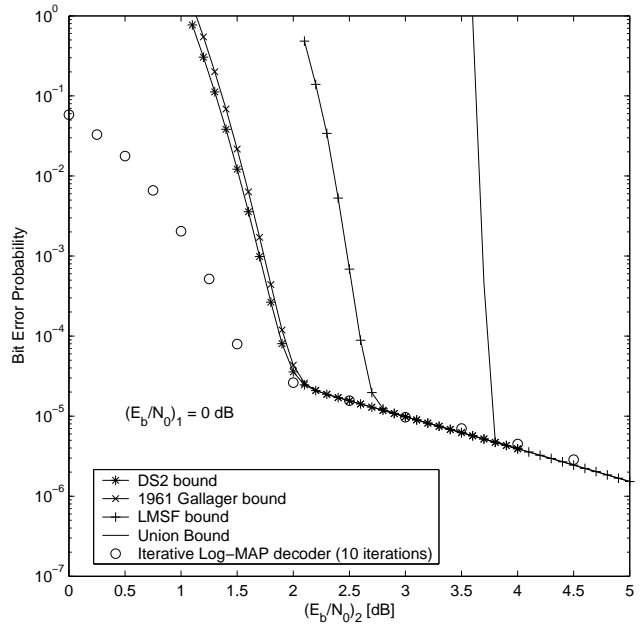

Fig. 2. Performance bounds for the bit error probability under ML decoding versus computer simulation results of iterative Log-MAP decoding (with 10 iterations).

### B. Attainable Channel Regions for Various Ensembles of Accumulate-Based Codes

In this section, we compare inner bounds on the attainable channel regions of accumulate-based codes under ML decoding. The comparison refers to three ensembles of rate one-third, as depicted in Fig. 3; the first is the ensemble of uniformly interleaved and non-systematic repeat-accumulate (RA) codes with $q = 3$ repetitions; the second and the third are uniformly interleaved and systematic ensembles of RA codes and accumulate-repeat-accumulate [1] (ARA) codes, respectively, where the number of repetitions is equal to $q = 6$ and, as a result of periodic puncturing, every third bit of the non-systematic part is transmitted (so the puncturing period is $p = 3$). For simplicity of notation, we make use of the abbreviations NSRA$(N, q)$, SPRA$(N, p, q)$ and SPARA$(N, M, p, q)$ for the encoders shown in Fig. 3 (a)–(c), respectively (i.e., the abbreviations 'NS' and 'SP' stand for 'non-systematic' and 'systematic and punctured', respectively). In this notation, $N$ is the input block length.

In [9], it is shown that the asymptotic growth rate of the distance spectra of these ensembles is such that the second, third, and fourth conditions of Theorem 3 are satisfied. The distance spectra of the SPRA and SPARA ensembles considered here were calculated in [9] using the techniques introduced in [1].

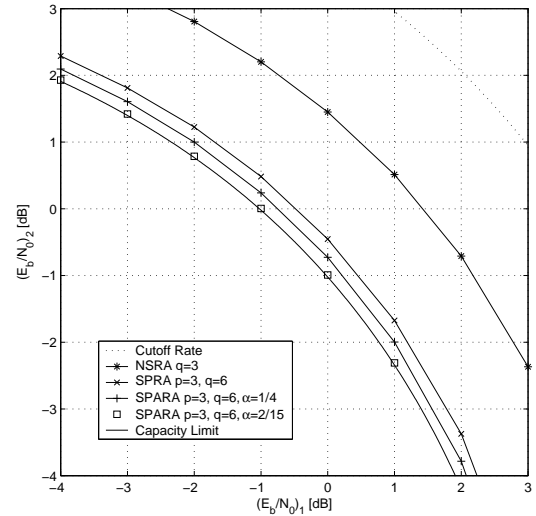

Fig. 4. Attainable channel regions for the rate 1/3 accumulate-based ensembles depicted in Fig. 3. The communication takes place over two parallel binary-input AWGN channels with $\alpha_1 = \alpha_2 = 0.5$. The capacity limit and the attainable channel region which refers to the cutoff rate are referenced.

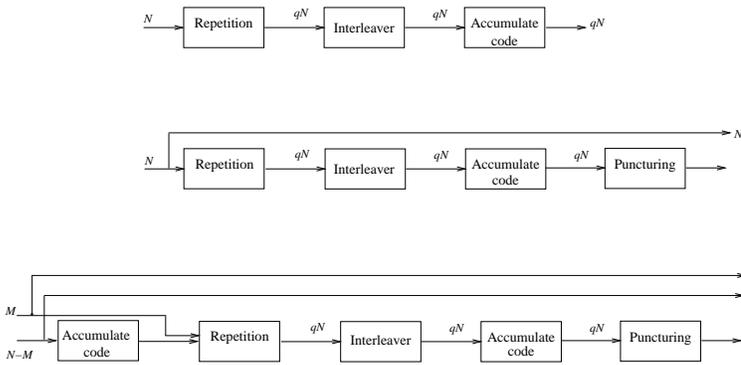

Fig. 3. Systematic and Non-systematic RA and ARA codes.

The improved performance of the ensembles of SPARA codes under ML decoding is demonstrated by the Gallager bounding technique (combined with the optimization of the tilting measures) in Fig. 4. In this figure, $\alpha \triangleq \frac{M}{3N}$ denotes the fraction of bits which do not pass through the outer accumulator of the code. This enlargement of the attainable channel region for the ensemble of SPARA codes is attributed to the distance spectral thinning effect [7] which is more pronounced for this ensemble as compared to the other two (see [9]). This in turn yields an improved inner bound on the attainable channel regions, as observed in Fig. 4. It is shown in this figure that for the SPARA ensemble with the parameters $p = 3, q = 6$ and $\alpha = \frac{2}{15}$, the gap between the inner bound on the attainable channel region under ML decoding and the capacity limit is less than 0.05 dB.

ACKNOWLEDGMENT

This research work was supported by a grant from Intel Israel and by the Taub and Shalom Foundations.

REFERENCES

[1] A. Abbasfar, K. Yao and D. Divsalar, "Maximum-likelihood decoding analysis of accumulate-repeat-accumulate codes," *Proceedings IEEE 2004 Global Telecommunications Conference (GLOBECOM 2004)*, pp. 514–519, 29 November–3 December, 2004, Dallas, Texas, USA.
[2] D. Divsalar, "A simple tight bound on error probability of block codes with application to turbo codes," *Telecommunications and Mission Operations (TMO) Progress Report* 42–139, JPL, pp. 1–35, November 15, 1999.
[3] T. M. Duman, *Turbo Codes and Turbo Coded Modulation Systems: Analysis and Performance Bounds*, Ph.D. dissertation, Elect. Comput. Eng. Dep., Northeastern University, Boston, MA, USA, May 1998.
[4] R. G. Gallager, *Low-Density Parity-Check Codes*, Cambridge, MA, USA, MIT Press, 1963.
[5] A. Khandekar, *Graph-based Codes and Iterative Decoding*, Ph.D. dissertation, California Institute of Technology, Pasadena, CA, USA, June 2002.
[6] R. Liu, P. Spasojevic and E. Soljanin, "Reliable channel regions for good binary codes transmitted over parallel channels," *IEEE Trans. on Information Theory*, vol. 52, pp. 1405–1424, April 2006.
[7] L. C. Perez, J. Seghers and D. J. Costello, "A distance spectrum interpretation of turbo codes," *IEEE Trans. on Information Theory*, vol. 42, pp. 1698–1709, November 1996.
[8] I. Sason and S. Shamai, "Performance analysis of linear codes under maximum-likelihood decoding: a tutorial," *Foundations and Trends in Communications and Information Theory*, vol. 3, no. 1–2, pp. 1–222, NOW Publishers, Delft, the Netherlands, July 2006.
[9] I. Sason and I. Goldenberg, "Coding for parallel channels: Gallager bounds for binary linear codes with applications to repeat-accumulate codes and variations," *submitted to IEEE Trans. on Information Theory*, June 2006. [Online]. Available: http://arxiv.org/abs/cs.IT/0607002.
[10] N. Shulman and M. Feder, "Random coding techniques for nonrandom codes," *IEEE Trans. on Information Theory*, vol. 45, no. 6, pp. 2101-2104, September 1999.